\documentclass[pra,showpacs,groupaddress,twocolumn,nofootinbib]{revtex4-1}

\usepackage{graphicx}
\usepackage{dcolumn}
\usepackage{amsmath}
\usepackage{epsfig}
\usepackage{bm}
\usepackage{bbold}
\usepackage{bbm}
\usepackage{amssymb}
\usepackage{hyperref}
\hypersetup{colorlinks=true,linkcolor=blue,citecolor=blue,urlcolor=black}

\begin{document}

\title{Quantum teleportation via maximum-confidence quantum measurements}

\author{L. Neves}

\author{M. A. Sol\'is-Prosser}
\email{msolisp@udec.cl}

\author{A. Delgado}

\affiliation{Center for Optics and Photonics, Universidad de Concepci\'on, Casilla 4016, Concepci\'on, Chile}
\affiliation{MSI-Nucleus on Advanced Optics, Universidad de Concepci\'on, Casilla 160-C, Concepci\'on, Chile}
\affiliation{Departamento de F\'isica, Universidad de Concepci\'on, Casilla 160-C, Concepci\'on, Chile}

\author{O. Jim\'{e}nez}
\affiliation{Departamento de F\'isica, Facultad de Ciencias B\'asicas, Universidad de Antofagasta, Casilla 170, Antofagasta, Chile}

\date{\today}

\begin{abstract}

We investigate the problem of teleporting unknown qudit states via pure quantum channels with nonmaximal Schmidt rank.  This process is mapped to the problem of discriminating among nonorthogonal symmetric states which are linearly dependent and equally likely. It is shown that by applying an optimized maximum-confidence (MC) measurement for accomplishing this task, one reaches the maximum possible teleportation fidelity after a conclusive event in the discrimination process, which in turn occurs with the maximum success probability. In this case, such fidelity depends only on the Schmidt rank of the channel and it is larger than the optimal one achieved, deterministically, by the standard teleportation protocol. Furthermore, we show that there are  quantum channels for which it is possible to apply a $k$-stage sequential MC measurement in the discrimination process such that a conclusive event at any stage leads to a teleportation fidelity above the aforementioned optimal one and, consequently, increases the overall success probability of teleportation with a fidelity above this limit. 

\end{abstract}

\date{\today}


\maketitle

\section{Introduction} 
\label{introduction}

Quantum teleportation is a quantum communication protocol where a sender, Alice, aims to transmit to a receiver, Bob, an unknown state of a $D$-dimensional quantum system (qudit) without sending the system itself. In the standard version of the protocol \cite{Bennett93}, Alice and Bob share two qudits in a pure and maximally entangled state (the quantum channel). She then performs a joint measurement on her half of the entangled pair and the system carrying the state to be teleported. After that, she communicates the outcome of her measurement to Bob through a classical channel with $2\log_2(D)$ bits capacity. Once Bob learns  Alice's outcome, he performs a unitary operation on his previously entangled system, which transforms its state into the original one that Alice aimed to transmit, and then the teleportation is complete. The consumption of a pure maximally entangled state and $2\log_2(D)$ bits of classical communication are both necessary and sufficient conditions for deterministic and faithful teleportation \cite{Bennett93}, which means that Alice transmits a quantum state to Bob with unit probability of success and unity fidelity.

When Alice and Bob share a pure but \emph{nonmaximally} two-qudit entangled state, deterministic \emph{and} faithful teleportation is no longer possible.\footnote{The only exception occurs when the quantum channel is a nonmaximally entangled  pure state of two $N$-dimensional systems with $N>D$ \protect\cite{Gour04}\protect. This case will not be addressed here.} The best they can do is to teleport in a optimal way according to some figure(s) of merit (e.g., average teleportation fidelity, probability of successful teleportation and classical communication cost). 
In this regard, many optimized schemes have been suggested to implement quantum teleportation \cite{Horodecki99,Vidal00,Banaszek00,Banaszek01,Laustsen02,Mor96,Li00,Bandyopadhyay00,Son01,Gu02,Roa03,Pati07}.  For instance, to obtain the optimal average teleportation fidelity deterministically, Alice and Bob have just to perform the standard protocol \cite{Vidal00,Banaszek00,Banaszek01}. On the other hand, if it is crucial for Bob to receive from Alice a perfect replica of an unknown qudit state, they can resort to the so-called \emph{perfect conclusive teleportation} \cite{Mor96}. In this scheme, faithful teleportation is possible with less than unity probability of success and the optimal protocol is the one that maximizes such probability. So far, several strategies have been proposed to implement the optimal protocol \cite{Mor96,Li00,Bandyopadhyay00,Son01,Gu02,Roa03}. In particular, for the purposes of the present work, we highlight the strategy of Ref.~\cite{Roa03} where it is shown that to accomplish perfect conclusive teleportation in a optimal way, Alice must implement an optimized measurement which discriminates unambiguously between a set of linearly \emph{independent} and equally likely symmetric states \cite{Ivanovic87,Chefles98-1,Chefles98-2,Jimenez07}.

Perfect conclusive teleportation applies only for a quantum channel with maximal Schmidt rank \cite{Horodecki99,Vidal00}\footnote{The Schmidt rank is the number of nonvanishing coefficients in the Schmidt decomposition of a pure bipartite quantum state. For two $D$-dimensional systems its maximum value is $D$. \label{FN:Schmidt}}. In this article we investigate the problem of teleporting via nonmaximally entangled states with \emph{nonmaximal} Schmidt rank. Since the success probability of faithful teleportation is zero in this case, we design an optimal \emph{imperfect conclusive teleportation} protocol which can be summarized as follows: Alice tries to teleport an unknown qudit state to Bob and succeeds with the maximum achievable probability. When she succeeds, the teleportation fidelity, although less than unit, is also the maximum achievable for their shared quantum channel. Such fidelity depends only on the Schmidt rank of this channel and it is larger than the optimal one achieved, deterministically, by the standard protocol \cite{Vidal00,Banaszek00,Banaszek01}. This is relevant, for instance, in a scenario where it is sufficient for Alice's and Bob's purposes to teleport with a fidelity above some threshold and their quantum channel allows this only in a probabilistic way. 

Similarly to the approach of Ref.~\cite{Roa03} for quantum channels with maximal Schmidt rank, we shall see here that if this number is not maximal, in the course of the teleportation process Alice is faced with the problem of discriminating among nonorthogonal symmetric states which are linearly \emph{dependent} and equally likely. Then, we show that by implementing an optimized maximum-confidence (MC) measurement \cite{Croke06} to accomplish this task, she can perform the optimal imperfect conclusive teleportation as described above. The MC discrimination for the aforementioned set of states was studied recently \cite{Jimenez11,Herzog12}. Specifically, in Ref.~\cite{Jimenez11} we have found the optimal positive operator valued measure (POVM) that maximizes our confidence in identifying each state in the set and
minimizes the probability of obtaining inconclusive results. We also showed that after an inconclusive result, the input states are mapped into a new set of equiprobable symmetric states within a lower-dimensional Hilbert space. If such space is not one dimensional, this new set could be submitted to another round of MC measurements, with which we could still gain some information about the input states, although
with less confidence than before. This process can be repeated in as many stages as allowed by the input states, and we called it \emph{sequential maximum-confidence} (SMC) measurements. 

Applying SMC measurements to our imperfect conclusive teleportation protocol, we shall see that the number of stages that Alice is allowed to implement, depends on both the Schmidt rank and Schmidt coefficients of the quantum channel, which she and Bob know in advance. We then show that there are quantum channels for which it is possible to implement a $k$-stage SMC measurement ($k>1$) such that conclusive events at any stage lead to a teleportation fidelity above the optimal one achieved by the standard protocol \cite{Vidal00,Banaszek00,Banaszek01}. In this case, there will be an increase of the overall probability of teleportation with a fidelity above that limit and, consequently, Alice and Bob will be able to save resources for accomplishing the protocol, as we shall discuss later. We shall as well discuss some side effects of our protocol, as for instance, the requirement of ancillary systems, the increase in the classical communication cost, and the reduction of the overall teleportation fidelity (which involves conclusive and inconclusive events) when compared with the one achieved by the optimal deterministic protocol.

The remainder of this article is organized as follows: In Sec.~\ref{review} we show how the teleportation process via nonmaximally entangled states can be mapped to the problem of discriminating nonorthogonal symmetric states. We then discuss the optimal deterministic and perfect conclusive protocols within this context. In Sec.~\ref{results1} we briefly review the discrimination of symmetric states via MC measurements and present the optimal imperfect conclusive teleportation protocol. An explicit example is described to illustrate it. In Sec.~\ref{sec:results2} we describe SMC measurements and  apply such a strategy to the teleportation process. Its benefits as well as its drawbacks are analyzed, and the previous example is used once again to illustrate the protocol. Finally, in Sec.~\ref{summary}, we summarize the results, discuss possible extensions of our work, and conclude.

\section{Quantum teleportation assisted by quantum state discrimination}
\label{review}

In this section we will describe the teleportation protocol within the framework of quantum computing \cite{Brassard98,NielsenBook}. Doing so, it will be straightforward to establish its connection with the problem of quantum state discrimination. In this regard we shall see that an optimal deterministic protocol and a probabilistic and faithful one can be associated with minimum-error and unambiguous discrimination strategies, respectively (reminding that the latter case is a known result \cite{Roa03}). This starting point will be important to understand how an optimal probabilistic and unfaithful teleportation protocol is connected with a maximum-confidence discrimination strategy, as we shall see in Sec.~\ref{results1}.

\subsection{Background}     \label{subsec:general}

Suppose that Alice wants to teleport to Bob an unknown pure state of a qudit given by 
\begin{equation}     \label{eq:unknown}
|\phi\rangle_3=\sum_{k=0}^{D-1}c_k|k\rangle_3,
\end{equation}
where $\sum_{k=0}^{D-1}|c_k|^2=1$ and the orthonormal set of states $\{|k\rangle_3\}$ forms the computational basis spanning a $D$-dimensional Hilbert space ${\cal H}_3$. The quantum channel to implement the protocol is given by a pure bipartite entangled state $|\Psi\rangle_{12}$, where system 1 belongs to Bob and system 2 to Alice. Without loss of generality, we assume that both systems live in a $D$-dimensional Hilbert space, ${\cal H}_1$ and ${\cal H}_2$, respectively. Using the Schmidt decomposition the state $|\Psi\rangle_{12}$ can be written as \cite{NielsenBook}
\begin{equation}
|\Psi\rangle_{12}=\sum_{m=0}^{N-1}a_m|m\rangle_{1}|m\rangle_{2},
\label{QUANTUMCHANNEL}	
\end{equation}
where $a_m$ are strictly positive real numbers satisfying  $\sum_{m=0}^{N-1}a_m^2=1$. For simplicity we also assume that these coefficients are decreasingly ordered, i.e., $a_i\geqslant a_{i+1}>0$. Note that we have included in the sum only the nonvanishing Schmidt coefficients, and thereby $N$ represents the Schmidt rank of the state $|\Psi\rangle_{12}$ (see footnote \ref{FN:Schmidt}), where $N\leqslant D$. As a final assumption, let the Schmidt basis, which is composed by $D$ orthonormal states $\{|m\rangle_i\}_{m=0}^{D-1}$,  coincide with the computational basis spanning ${\cal H}_i$ for $i=1,2$. If this is not the case in the beginning, Alice and Bob can apply local unitary rotations to align the Schmidt basis with the computational one \cite{Vidal00}. 

As we mentioned in the Introduction, Alice must now perform a joint measurement on the quantum systems in her possession, in this case systems 2 and 3. Generalizing the approach of Refs.~\cite{Brassard98,NielsenBook} for qubits to higher-dimensional spaces, this measurement can be described by a two-step process. In the first step, Alice implements a unitary operation given by a generalized controlled-\textsc{not} gate, $\hat{G}^\text{\sc xor}_{23}$, having system 2 as the control qudit and system 3 as the target qudit (see Fig.~\ref{fig:Q_circuit1}). The action of $\hat{G}^\text{\sc xor}_{23}$ onto the computational basis is defined by \cite{Alber01,Jex03}
\begin{equation}    \label{eq:GXOR}
\hat{G}^\text{\sc xor}_{23}|i\rangle_2|j\rangle_3=|i\rangle_2|i\ominus j\rangle_3,
\end{equation}
where $\ominus$ denotes subtraction modulo $D$. Applying this operator on the initial three-system state $|\Psi\rangle_{12}|\phi\rangle_3$ and using Eqs.~(\ref{eq:unknown})--(\ref{eq:GXOR}), we obtain
\begin{equation}
\hat{G}^\text{\sc xor}_{23}|\Psi\rangle_{12}|\phi\rangle_3=
\frac{1}{D}\sum_{l,k=0}^{D-1}\hat{Z}_1^{D-l}\hat{X}_1^{k}|\phi\rangle_1|\nu_l\rangle_2|k\rangle_3,
\label{BASE}
\end{equation}
where $\hat{Z}$ and $\hat{X}$ are the generalized Pauli operators, defined by their action on the computational basis as 
\begin{equation}     \label{eq:Z}
\hat{Z}|m\rangle=e^{2\pi im/D}|m\rangle,
\end{equation}
and
\begin{equation}
\hat{X}|m\rangle=|m\oplus 1\rangle,
\end{equation}
respectively, where $\oplus$ denotes addition modulo $D$. The states $|\nu_l\rangle_2$ in Eq.~(\ref{BASE}) are given by 
\begin{equation}
|\nu_l\rangle_2=\hat{Z}_2^l\sum_{k=0}^{N-1}a_k|k\rangle_2,
\label{SYMMETRICSTATES}
\end{equation}
for $l=0,\dots,D-1$. They form a set of $D$ symmetric states\footnote{A set of states is symmetric under the action of a unitary $\hat{Z}$, if they satisfy the following conditions:  $|\nu_l\rangle=\hat{Z}|\nu_{l-1}\rangle=\hat{Z}^{l}|\nu_0\rangle$, $|\nu_0\rangle=\hat{Z}|\nu_{D-1}\rangle$ \cite{Chefles98-1}. This is the case for the states in Eq.~(\ref{SYMMETRICSTATES}) with the unitary given by Eq.~(\ref{eq:Z}).} under the action of $\hat{Z}_2$, and are characterized by the state coefficients $a_k$ defining the quantum channel in Eq.~(\ref{QUANTUMCHANNEL}). Additionally, the Schmidt rank of this channel, $N$, will determine if the set is linearly independent ($N=D$) or linearly dependent ($N<D$).

The second step of Alice's procedure is to perform measurements on the systems 2 and 3 individually. From Eq.~(\ref{BASE}) we see that the state of system 3 can be perfectly determined by a projective measurement in the computational basis. The measurement of system 2, however, will depend on the features of the quantum channel [which defines the features of the set $\{|\nu_l\rangle_2\}$ in Eq.~(\ref{SYMMETRICSTATES})] and a prearrangement between Alice and Bob regarding which type of teleportation protocol they want (or they can) implement. Therefore, the teleportation process is mapped to the problem of Alice's ability to perform an optimized measurement which discriminates $D$ symmetric states $\{|\nu_l\rangle_2\}$ occurring with equal prior probabilities $1/D$. In what follows we discuss the possibilities for teleportation in the context of state discrimination.

\subsection{Perfect teleportation}

\begin{figure}[t]
\centerline{
\includegraphics[width=.48\textwidth]{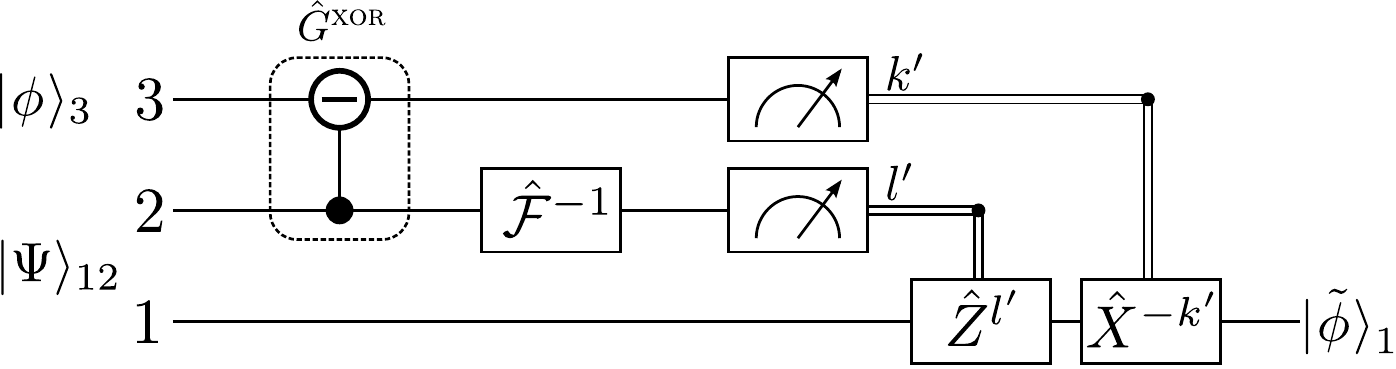}}
\caption{Quantum circuit for standard teleportation. For $|\Psi\rangle_{12}$ maximally entangled, $|\tilde\phi\rangle\equiv|\phi\rangle$ and the teleportation fidelity is 1. Otherwise, $|\tilde\phi\rangle\neq|\phi\rangle$ and the maximum fidelity is the one given by Eq.~(\ref{eq:F_opt}). }
\label{fig:Q_circuit1} 
\end{figure}

In order to achieve faithful and deterministic teleportation, Alice and Bob must share a maximally entangled state in the first place. When this is the case, we have in Eq.~(\ref{QUANTUMCHANNEL}) $N=D$ and $a_m=1/\sqrt{D}$ for all $m$. Thus, the symmetric states in Eq.~(\ref{SYMMETRICSTATES}) become 
\begin{equation}
|\nu_l\rangle_2=\hat{\mathcal{F}}_2|l\rangle_2,
\end{equation}
where $\hat{\mathcal{F}}$ is the discrete Fourier transform defined by
\begin{equation}
\hat{\mathcal{F}}=\frac{1}{\sqrt{D}}\sum_{m,n=0}^{D-1}e^{2\pi imn/D}|m\rangle\langle n|.
\label{eq:Fourier}
\end{equation}
Therefore, system 2 is mapped to a set of orthogonal states which can be perfectly and deterministically discriminated. Due to this fact, Alice can perform faithful and deterministic teleportation. The whole process can be understood from Eq.~(\ref{BASE}) and the quantum circuit shown in Fig.~\ref{fig:Q_circuit1}. Alice applies a $\hat{G}^\text{\sc xor}_{23}$ between systems 2 and 3, followed by an inverse Fourier transform at 2. Then she measures both systems in the computational basis\footnote{This two-step process is equivalent to performing the complete Bell-state measurement \cite{Brassard98,NielsenBook}.} and sends a $2\log_2(D)$-bit message to Bob through a classical channel communicating the result $(l',k')$ that she has obtained. Finally, Bob uses this information and apply the unitary transformations $\hat{X}_1^{-k'}\hat{Z}_1^{l'}$ on his system whose state becomes a perfect replica of the unknown state $|\phi\rangle$.

\subsection{Teleportation and singlet fraction}

Before proceeding, we briefly recall a result that will be useful for the discussions that follow. In Ref.~\cite{Horodecki99}, Horodecki \emph{et al.} demonstrated that for any given initial entangled state $\hat{\rho}_{12}$ (pure or mixed) shared between Alice and Bob, the optimal teleportation fidelity over all possible protocols performed via local operations an classical communication (LOCC) is given by 
\begin{equation}     \label{eq:Horodecki}
F^{\mathcal{S}}=\frac{Df^{\mathcal{S}}+1}{D+1}.
\end{equation}
Here, the superscript $\mathcal{S}$ indicates whether the protocol belongs to a class of deterministic or probabilistic protocols. The parameter $f^\mathcal{S}$,  called the \emph{singlet fraction}, is defined as 
\begin{equation}     \label{eq:Singlet-fraction}
f^\mathcal{S}=\max_{\tilde\Psi}\left[{}_{12}\langle\tilde\Psi|\hat{\rho}_{12}|\tilde\Psi\rangle_{12}\right],
\end{equation}
and measures the maximum fidelity that can be achieved between $\hat{\rho}_{12}$ and a maximally entangled state $|\tilde\Psi\rangle_{12}$ when the former is subjected to $\mathcal{S}$ transformations using LOCC. In the case of faithful and deterministic teleportation discussed above, we have that the singlet fraction will be unit, and so will be the fidelity in (\ref{eq:Horodecki}).

In the remainder of this work we shall associate $\mathcal{S}$ not only with the class of the protocol but also with a given state discrimination strategy within that class. In addition, we will see that the singlet fraction is equal to the confidence achieved in the discrimination of the symmetric states in Eq.~(\ref{SYMMETRICSTATES}) when such a strategy is employed.

\subsection{Optimal deterministic teleportation}     
\label{subsec:ME_telep}

Let us now assume that Alice and Bob share a nonmaximally entangled pure state and they agree beforehand to implement teleportation deterministically.
Following the discussion in Sec.~\ref{subsec:general}, after applying the $\hat{G}^\text{\sc xor}_{23}$, system 2 is mapped to a set of $D$ equally likely nonorthogonal symmetric states given by Eq.~(\ref{SYMMETRICSTATES}). These states cannot be discriminated, simultaneously, in a perfect and deterministic way. Due to her prior agreement with Bob, Alice must then abandon the requirement for perfection and implement a measurement with $D$ possible outcomes where errors are allowed to occur. In this case, the optimized measurement will be the one that minimizes the probability of making an erroneous identification of the state, which is known as minimum-error (ME) discrimination \cite{Helstrom76,Holevo73}. It turns out that for the aforementioned set of states of system 2, the physical implementation of the optimal ME measurement follows exactly the same procedure of the standard teleportation protocol (see Fig.~\ref{fig:Q_circuit1}), that is, an inverse Fourier transform followed by a projective measurement on the computational basis \cite{Ban97,Jimenez07,Jimenez11}\footnote{The physical implementation of an optimal ME measurement that discriminates among $D$ equally likely symmetric states defined by Eq.~(\ref{SYMMETRICSTATES}) is given by the $D$-outcome projective measurement $\{\mathcal{F}|l\rangle\langle l|\mathcal{F}^{-1}\}_{l=0}^{D-1}$. It applies for both linearly independent and linearly dependent states. In the latter case, this measurement realizes the optimal POVM on the lower-dimensional subspace where the symmetric states are restricted in \cite{Jimenez11}. \label{FN:opt_ME} }. Therefore, when Alice applies the $\hat{G}^\text{\sc xor}_{23}$, measures system 3 in the computational basis, and implements an optimized ME measurement on system 2, she is just performing the standard teleportation (which is completed as indicated in the quantum circuit of Fig.~\ref{fig:Q_circuit1}). By performing standard teleportation, it is guaranteed by Refs. \cite{Vidal00,Banaszek00} that the fidelity between Bob's final state and Alice's unknown state, averaged over all possible input states $|\phi\rangle$, will be the maximum achievable 
\begin{equation}     \label{eq:F_opt}
\frac{1}{D+1}\left[1+\left(\sum_{m=0}^{N-1} a_m\right)^2\right]\equiv F^{\rm ME},
\end{equation}
where $a_m$ are the Schmidt coefficients of the entangled state given by Eq.~(\ref{QUANTUMCHANNEL}). The superscript ``ME'' now indicates the optimal discrimination strategy adopted by Alice to achieve this optimal fidelity in a deterministic protocol. 
Comparing Eqs.~(\ref{eq:Horodecki}) and (\ref{eq:F_opt}) we find that 
\begin{equation}     \label{eq:Sing_frac_ME}
f^{\rm ME}=\frac{1}{D}\left(\sum_{m=0}^{N-1} a_m\right)^2.
\end{equation}
The quantity on the right-hand side is precisely the probability of correctly discriminating among equally likely symmetric states via ME measurements, $P_{\rm corr}^{\rm ME}$, which in turn is equal to the maximum confidence achieved by a ME measurement in such discrimination,\footnote{Here, the conditional probability $[P(\nu_l|l)]^{\mathcal{S}}$ will represent our \emph{maximum} confidence in taking a measurement outcome $l$ to indicate the input state as $|\nu_l\rangle$, when the discrimination strategy $\mathcal{S}$ is employed. \label{FN:MC-def}} $[P(\nu_l|l)]^{\rm ME}$ \cite{Jimenez11}, that is, 
\begin{equation}     \label{eq:sing_frac_ME}
f^{\rm ME} = P_{\rm corr}^{\rm ME} = [P(\nu_l|l)]^{\rm ME}.
\end{equation}

The connection between optimal deterministic teleportation and optimal ME discrimination established here, will be important to understand how the optimal teleportation via MC measurements works in practice, as we will see in Sec.~\ref{results1}.

\subsection{Optimal perfect conclusive teleportation}
\label{subsec:UD_telep}

From the discussion in Sec.~\ref{subsec:general}, it follows that in order to teleport faithfully via a nonmaximally entangled channel, Alice must be able to discriminate $D$ nonorthogonal states of system 2 [$|\nu_l\rangle_2$ in Eq.~(\ref{SYMMETRICSTATES})] without error. This is possible only if inconclusive outcomes are allowed to occur \cite{Ivanovic87} and the states are linearly independent \cite{Chefles98-2}, that is, $N=D$ in Eq.~(\ref{SYMMETRICSTATES}). The first condition makes the protocol probabilistic, while the second requires the Schmidt rank of the channel to be maximal. Assuming that the latter condition holds, the optimal teleportation protocol will be the one in which Alice's measurement strategy to determine  the states $|\nu_l\rangle_2$ minimizes (maximizes) the probability of inconclusive (conclusive) outcomes. This strategy is known as optimal unambiguous discrimination (UD), and the optimized measurement that implements it for equally likely symmetric states is known \cite{Chefles98-1,Jimenez07}. 

The teleportation protocol via optimal UD of symmetric states has been studied in Ref.~\cite{Roa03}. It is a perfect conclusive protocol because after a conclusive event in Alice's discrimination process, the teleportation proceeds as in the standard scheme and in the end Bob receives a perfect replica of the unknown state. Thus, the teleportation fidelity will be
\begin{equation}
F^{\rm UD}_{\rm s} = 1,
\end{equation}
where the subscript ``s'' labels conclusive (or successful) events. In terms of the singlet fraction, we have from Eq.~(\ref{eq:Horodecki})
\begin{equation}
f^{\rm UD}_{\rm s}=1 = [P(\nu_l|l)]^{\rm UD}_{\rm s},
\end{equation}
where we used the fact that the confidence (see footnote \ref{FN:MC-def}) achieved in the UD strategy is unity for each conclusive event.

After an inconclusive event, unambiguous discrimination is forbidden \cite{Ivanovic87,Chefles98-2,Chefles98-1} and, accordingly, perfect teleportation. When it happens, Alice and Bob just discard the attempt and start the process again with another copy of the shared entangled pair, until she succeeds. In terms of resources this protocol will be more expensive than a deterministic one since it will require the use of ancillary systems (ancilla), one extra bit of classical communication for each teleportation attempt until the successful one and, in general, more copies of the quantum channel to be efficient. The quantum circuits for this protocol are shown in Figs.~\ref{fig:Q_circuit2}(a) and \ref{fig:Q_circuit2}(b), and they will be better understood along the discussion of teleportation via MC measurements.

\section{Quantum Teleportation via maximum-confidence measurements}
\label{results1}

When Alice and Bob share a quantum channel with nonmaximal Schmidt rank, i.e., $N<D$ in Eq.~(\ref{QUANTUMCHANNEL}), the states $|\nu_l\rangle_2$ in Eq.~(\ref{SYMMETRICSTATES}) become linearly dependent. In this case, perfect conclusive teleportation described above is impossible. It is possible, however, to implement optimal \emph{imperfect} conclusive teleportation. As we shall see in this section, this is achieved when Alice applies MC measurements \cite{Croke06,Jimenez11} to discriminate the states $|\nu_l\rangle_2$. With such strategy, conclusive events in the discrimination process occur with maximum success probability and lead to a teleportation with the maximum achievable fidelity for the quantum channel. We show that this fidelity depends only on the Schmidt rank of the channel and, although less than a unit, it is larger  than the one achieved deterministically [see Eq.~(\ref{eq:F_opt})]. In the following we shall, first, briefly review MC discrimination for symmetric states \cite{Jimenez11}, and then apply this strategy to the teleportation process. The advantages as well as the drawbacks of the protocol are discussed and an example is presented to illustrate it. It is worth mentioning that the problem addressed here could equivalently be thought as the teleportation of $D$-dimensional qudit states using two entangled $N$-dimensional qudits. In fact, this problem has been discussed in Ref.~\cite{Laustsen02} from a completely different approach and restricted to maximally entangled states only. In our protocol, however, this is just a special case that occurs when all the Schmidt coefficients in (\ref{QUANTUMCHANNEL}) are equal as will be shown later.

\subsection{Optimized maximum-confidence discrimination of symmetric pure states}
\label{subsec:MC-review}

In the problem of quantum state discrimination, the MC strategy is implemented through an optimized measurement in which each conclusive outcome leads us to identify each state with the maximum possible confidence, while keeping at the minimum the probability of obtaining inconclusive results \cite{Croke06}. In fact, the MC strategy interpolates between ME and UD strategies: When the confidence is the same for all states and there is no inconclusive result, MC and ME coincide; when the confidence is equal to unity for
each conclusive result MC corresponds to UD strategy.

Recently, we have addressed the problem of discriminating with maximum confidence among $D$ symmetric states of $N$-dimensional qudits ($N<D$), prepared with equal prior probabilities $1/D$. For this linearly dependent set, defined in Eq.~(\ref{SYMMETRICSTATES}) for $N<D$, we found the optimal POVM and determined its physical implementation \cite{Jimenez11}. The latter is achieved by the following procedure: First, a two-dimensional ancilla, initially prepared in the logical state $|0\rangle_{\rm a}$, is coupled with the qudit (for convenience labeled as system 2) through a unitary operation $\hat{U}_{2\rm a}$ acting on the tensor product Hilbert space $\mathcal{H}_2\otimes\mathcal{H}_{\rm a}$. The optimal unitary, derived in Ref. \cite{Jimenez11}, transforms the input states $|\nu_l\rangle_2$ as follows:
\begin{equation}
\hat U_{2{\rm a}}|\nu_l\rangle_2|0\rangle_{\rm a}=\sqrt{1-P^?_\mathbbm{1}}|u_l\rangle_2|0\rangle_{\rm a}+\sqrt{P^?_\mathbbm{1}}|\xi_l\rangle_2|1\rangle_{\rm a},
\label{MC}
\end{equation}
where $P^?_\mathbbm{1}$ (the role of the subscript will become clear later)  is the minimum probability of obtaining an inconclusive result, given by \cite{Jimenez11}
\begin{equation}     \label{eq:min_P?}
P^?_\mathbbm{1}=1-Na_{N-1}^2,
\end{equation}
where $a_{N-1}$ is the minimum coefficient for the states $|\nu_l\rangle_2$ in Eq.~(\ref{SYMMETRICSTATES}). The normalized states $|u_l\rangle_2$ and $|\xi_l\rangle_2$ can be written as
\begin{equation}
|u_l\rangle_2=\hat{Z}_2^l\sum_{k=0}^{N-1}\frac{|k\rangle_2}{\sqrt{N}}
\label{MCU}
\end{equation}
and 
\begin{equation}
|\xi_l\rangle_2=\hat{Z}_2^l\sum_{k=0}^{N-1}\sqrt{\frac{a^2_k-a^2_{N-1}}{P^?_\mathbbm{1}}}|k\rangle_2,
\label{MCXI}
\end{equation}
respectively, with $\hat{Z}$ defined in Eq.~(\ref{eq:Z}). Note that those states are still symmetric and equally likely. Each $|u_l\rangle_2$ ($|\xi_l\rangle_2$) occurs with probability $1-P^?_\mathbbm{1}$ ($P^?_\mathbbm{1}$).

After the unitary coupling, the ancilla is measured on the computational basis, namely $\{|0\rangle_{\rm a},|1\rangle_{\rm a}\}$. If it is projected on $|0\rangle_{\rm a}$, the initial states are mapped into $|u_l\rangle_2$. This outcome is interpreted as conclusive (or successful) in the sense that the transformed states can now be subjected to a measurement that will discriminate them---and hence discriminate $|\nu_l\rangle_2$---with the maximum achievable confidence. Otherwise, if the ancilla is projected on $|1\rangle_{\rm a}$, the initial states are mapped into $|\xi_l\rangle_2$, and this result is inconclusive (or a failure) in the sense that there is no measurement that discriminates them with the maximum achievable confidence. When the result is conclusive, we have shown in Ref. \cite{Jimenez11} that the MC and ME strategies coincide, and so the optimized measurement for both is the same. As we discussed in Sec.~\ref{subsec:ME_telep} (see also footnote \ref{FN:opt_ME}), this measurement is performed by, first, applying an inverse Fourier transform [see Eq.~(\ref{eq:Fourier})] acting on a $D$-dimensional Hilbert space, and then projecting on the computational basis that spans this space. Doing so, our confidence in taking an outcome $l$ to identify the input state as $|\nu_l\rangle_2$ will be\footnote{To fix notation, the subscripts ``$j,\mathbbm{k}$'' (for $j=$ ``s'' or ``f'' and $\mathbbm{k}$ a positive integer) indicate that the quantity (or operator) before them is related to a conclusive ($j\rm =s$) or inconclusive ($j\rm =f$) event at the $\mathbbm{k}$th stage of a possible sequential MC measurement, as will be discussed in Sec.~\ref{sec:results2}. \label{FN:Notation} }
\begin{equation}    \label{eq:MC_probability}
[P(\nu_l|l)]_{{\rm s},\mathbbm{1}}^{\rm MC}=\frac{N}{D},
\end{equation}
which is the maximum achievable for equally likely symmetric pure states \cite{Jimenez11,Herzog12}.

\subsection{Optimal imperfect conclusive teleportation}

\begin{figure}[t]
(a)\\[3mm]
\centerline{ 
\includegraphics[width=.45\textwidth]{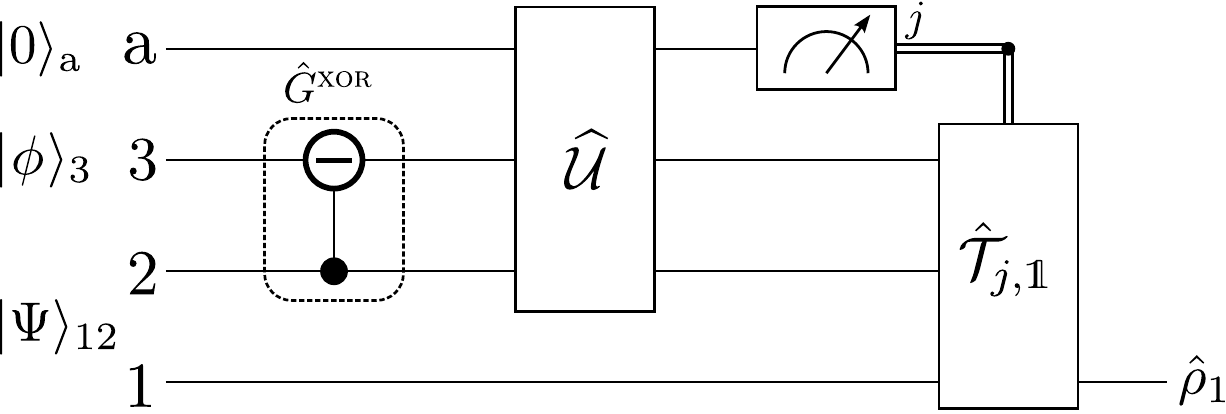}}
\vskip .5cm
(b)\\[3mm]
\centerline{
\includegraphics[width=.48\textwidth]{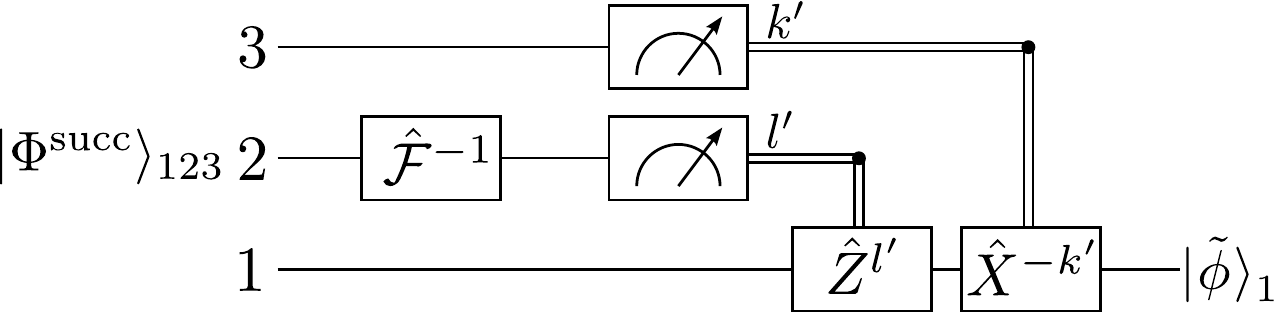}}
\vskip .5cm
\caption{Quantum circuits for optimal conclusive teleportation: Perfect when $N=D$ and imperfect when $N<D$. (a) Here $\hat{\mathcal{U}}=\hat{U}_{\rm 2a}\otimes\hat{I}_3$, where $\hat{I}_3$ is the identity on the Hilbert space of system 3 and the action of $\hat{U}_{\rm 2a}$ is defined by Eq.~(\ref{eq:MC_transform}). The box $\hat{\mathcal{T}}_{j,\mathbbm{1}}$ represents the quantum circuit that completes the protocol after the outcome $j$ in the measurement of the ancilla by Alice is communicated to Bob. $\hat{\rho}_1$ is the final state of Bob's system which depends on operation $\hat{\mathcal{T}}_{j,\mathbbm{1}}$. (b) In case of conclusive events, i.e., $j=\rm s$ the teleportation is accomplished as in the standard scheme (see Fig.~\ref{fig:Q_circuit1}).  The final state of system 1 will be either a faithful replica of the unknown state $|\phi\rangle$ (perfect protocol) or an unfaithful one with the optimal fidelity given by Eq.~(\ref{STRATEGY-CONCLUSIVE}) (imperfect protocol). \label{fig:Q_circuit2} }
\end{figure}

We show now that by applying the MC strategy described above, Alice and Bob can perform optimal imperfect conclusive teleportation. Following the protocol from the beginning, Alice first applies the $\hat{G}^\text{\sc xor}_{23}$ between systems 2 and 3 in her possession. She then adds a two-dimensional ancilla prepared in the logical state $|0\rangle_{\rm a}$ and applies the unitary transformation $\hat{U}_{2\rm a}$ [defined in Eq.~(\ref{MC})] between this ancilla and system 2. Using Eqs.~(\ref{BASE}) and (\ref{MC}) it can be shown that the four-system state can be written as 
\begin{eqnarray}
\hat U_{2\rm a}\hat{G}^\text{\sc xor}_{23}|\Psi\rangle_{12}|\phi\rangle_3|0\rangle_{\rm a} &=& \sqrt{1-P^?_\mathbbm{1}}|\Phi^{\rm succ}\rangle_{123}|0\rangle_{\rm a}
\nonumber\\[2mm]
& & \text{}+\sqrt{P^?_\mathbbm{1}}|\Phi^{\rm fail}\rangle_{123}|1\rangle_{\rm a},
\label{eq:MC_transform}
\end{eqnarray}
where $P^?_\mathbbm{1}$ is given by Eq.~(\ref{eq:min_P?}) and $|\Phi^{\rm succ}\rangle_{123}$ ($|\Phi^{\rm fail}\rangle_{123}$) denotes the normalized three-system state after a conclusive (an inconclusive) outcome in the measurement of the ancilla. They are given, respectively, by 
\begin{equation}
|\Phi^{\rm succ}\rangle_{123}=\frac{1}{D}\sum_{l,k=0}^{D-1}\hat Z_1^{D-l}\hat X_1^{k}|\phi\rangle_1|u_l\rangle_2|k\rangle_3,
\label{S1}
\end{equation}
where $|u_l\rangle_2$ is defined by Eq.~(\ref{MCU}), and   
\begin{equation}
|\Phi^{\rm fail}\rangle_{123}=\frac{1}{D}\sum_{l,k=0}^{D-1}\hat Z_1^{D-l}\hat X_1^{k}|\phi\rangle_1|\xi_l\rangle_2|k\rangle_3.
\label{S2}
\end{equation}
with $|\xi_l\rangle_2$ defined by Eq.~(\ref{MCXI}). Now, Alice measures the ancilla on the computational basis and communicates her outcome to Bob by sending him one bit of information through a classical channel. The process so far is depicted in the quantum circuit of Fig.~\ref{fig:Q_circuit2}(a). Each one of the two possible Alice's outcomes leads to one of the three-system states above, and each of them allows her to complete the teleportation process with different fidelities as we shall discuss next. In the circuit of Fig.~\ref{fig:Q_circuit2}(a) this is represented by the box $\hat{\mathcal{T}}_{j,\mathbbm{1}}$ (see footnote \ref{FN:Notation}).

\subsubsection{Conclusive events}
\label{subsec:Conc_telep}

A conclusive outcome in the measurement of the ancilla makes the transformation $|\Psi\rangle_{12}|\phi\rangle_3\rightarrow|\Phi^{\rm succ}\rangle_{123}$ [see Eq.~(\ref{eq:MC_transform})] or, equivalently, $|\nu_l\rangle_2\rightarrow|u_l\rangle_2$ [see Eqs.~(\ref{SYMMETRICSTATES}) and (\ref{MCU}), respectively]. From Eq.~(\ref{eq:min_P?}), such event occurs with the maximum success probability  $1-P^?_\mathbbm{1}=Na_{N-1}^2$. Alice now proceeds with the teleportation process by measuring systems 2 and 3. As usual, system 3 is subjected to a projective measurement on the computational basis. On the other hand, system 2 is subjected to the final measurement that will identify each state $|u_l\rangle_2$---and accordingly each $|\nu_l\rangle_2$---with maximum confidence. As described above and shown in Ref.~\cite{Jimenez11}, this final measurement is exactly the same as the optimized measurement that discriminates among equally likely symmetric states with the minimum probability of error (see footnote~\ref{FN:opt_ME}). Therefore, from the discussion in Sec.~\ref{subsec:ME_telep}, this implies that Alice performs the same measurement of the standard teleportation protocol, and, doing this, it can be ensured by Refs.~\cite{Vidal00,Banaszek00} that she will teleport the unknown state to Bob with the \emph{maximum} achievable fidelity. This fidelity, denoted by $F^{\rm MC}_{{\rm s},\mathbbm{1}}$ (see footnote \ref{FN:Notation}), will be given by Eq.~(\ref{eq:F_opt}) taking into account that $|\nu_l\rangle_2\rightarrow|u_l\rangle_2 \Rightarrow a_m\rightarrow N^{-1/2}$ for all $m$. Thus,
\begin{equation}
F^{\rm MC}_{{\rm s},\mathbbm{1}}=\frac{N+1}{D+1}.
\label{STRATEGY-CONCLUSIVE}
\end{equation}
The complete quantum circuit implementing this protocol is obtained by inserting the circuit of Fig.~\ref{fig:Q_circuit2}(b) into the box $\hat{\mathcal{T}}_{j,\mathbbm{1}}$ in Fig.~\ref{fig:Q_circuit2}(a) with $j={\rm s}$.

The teleportation fidelity in (\ref{STRATEGY-CONCLUSIVE}) depends only on the Schmidt rank, $N$, of the quantum channel unlike the deterministic protocol whose fidelity depends on both the Schmidt rank and Schmidt coefficients of the channel [see Eq.~(\ref{eq:F_opt})]. Consequently, for a given $N$, the fidelity achieved by teleporting via MC measurements will be the same for any entangled state shared between Alice and Bob. The success probability $Na_{N-1}^2$, on the other hand, depends on the particular quantum channel. In order to compare the fidelities for each protocol, we first find the singlet fraction for the present case. Comparing Eqs.~(\ref{eq:Horodecki}) and (\ref{STRATEGY-CONCLUSIVE}) and using Eq.~(\ref{eq:MC_probability}) we have 
\begin{equation}
f^{\rm MC}_{{\rm s},\mathbbm{1}}=\frac{N}{D}=[P(\nu_l|l)]_{{\rm s},\mathbbm{1}}^{\rm MC}.
\end{equation}
Since $[P(\nu_l|l)]_{{\rm s},\mathbbm{1}}^{\rm MC}\geqslant [P(\nu_l|l)]^{\rm ME}$ \cite{Jimenez11}, and so $f^{\rm MC}_{{\rm s},\mathbbm{1}}\geqslant f^{\rm ME}$ we finally obtain
\begin{equation}     \label{eq:Fmc-vs-Fme}
F^{\rm MC}_{{\rm s},\mathbbm{1}}\geqslant F^{\rm ME}.
\end{equation}
Therefore, if Alice applies the MC strategy instead of ME to discriminate the states $|\nu_l\rangle_2$ given by Eq.~(\ref{SYMMETRICSTATES}), after a conclusive result she will always teleport with a fidelity larger than the optimal one achieved, deterministically, by standard teleportation. The equality in (\ref{eq:Fmc-vs-Fme}) holds only when the entangled state in Eq.~(\ref{QUANTUMCHANNEL})---and hence $|\nu_l\rangle_2$ in (\ref{SYMMETRICSTATES})---has $a_m=N^{-1/2}$ for all $m$, since in this case MC and ME strategies coincide, as discussed above. (The fidelity for this particular channel has been derived in Ref. \cite{Laustsen02} using a different approach.) 
The price to pay to teleport with better fidelity is that the protocol becomes probabilistic and the required resources increase in the same way as in the perfect conclusive protocol described in Sec.~\ref{subsec:UD_telep}. 

Two extreme cases for the conclusive teleportation fidelity can be seen from Eq.~(\ref{STRATEGY-CONCLUSIVE}). First, it reaches its minimum value when $N=1$, that is, when the quantum channel in Eq.~(\ref{QUANTUMCHANNEL}) is separable. This value is the maximal fidelity of teleportation via classical channel \cite{Horodecki99} 
\begin{equation}     \label{eq:F_clas}
F^{\rm clas}=\frac{2}{D+1}.
\end{equation}
Consequently, as long as $N>1$, the conclusive teleportation fidelity $F^{\rm MC}_{{\rm s},\mathbbm{1}}$ will always exceed $F^{\rm clas}$ by an amount of $(N-1)/(D+1)$. In the second case, the fidelity reaches its maximum value of 1 when $N=D$, that is, when the quantum channel in Eq.~(\ref{QUANTUMCHANNEL}) has maximal Schmidt rank, so the imperfect conclusive protocol becomes perfect. 
In fact, regarding the success probability and the conclusive teleportation fidelity, the optimal imperfect conclusive protocol presented here interpolates between the optimal deterministic unfaithful ($a_m=N^{-1/2}, \forall\ m$) and the perfect conclusive ($N=D$) protocols described in Sec.~\ref{subsec:ME_telep} and Sec.~\ref{subsec:UD_telep}, respectively. This is so because,  as we mentioned earlier, MC strategy interpolates between ME and UD strategies when the above conditions are accomplished.

\subsubsection{Inconclusive events}
\label{subsub:telep_inc}

An inconclusive outcome in the measurement of the ancilla makes the transformation $|\Psi\rangle_{12}|\phi\rangle_3\rightarrow|\Phi^{\rm fail}\rangle_{123}$ [see Eq.~(\ref{eq:MC_transform})] or, equivalently, $|\nu_l\rangle_2\rightarrow|\xi_l\rangle_2$ [see Eqs.~(\ref{SYMMETRICSTATES}) and (\ref{MCXI}), respectively]. Such event occurs with the minimum failure probability given by Eq.~(\ref{eq:min_P?}). Since the $D$ states $|\xi_l\rangle_2$ form a new set of equally likely symmetric states restricted to a subspace of an $N$-dimensional space, they could be subjected to a new round of discrimination. If so, the teleportation process could be completed, although the achieved fidelity would be even lower than the conclusive one, no matter the strategy adopted between Alice and Bob. 

Let us assume here that in consequence of a prearrangement with Bob, Alice applies the ME strategy to discriminate the states $|\xi_l\rangle_2$, after an inconclusive result in the MC measurement. From our discussion in Sec.~\ref{subsec:ME_telep} and using Eq.~(\ref{MCXI}), the fidelity of the teleported state after the protocol is completed will be given by Eq.~(\ref{eq:F_opt}) with $ a_m\rightarrow \sqrt{(a_m^2-a_{N-1}^2)/P_\mathbbm{1}^?}$ for all $m$. Denoting this fidelity by $F^{\rm ME}_{{\rm f},\mathbbm{1}}$ (see footnote \ref{FN:Notation}) we obtain  
\begin{eqnarray}
F^{\rm ME}_{{\rm f},\mathbbm{1}}&=&
F^{\rm clas} + \sum_{\stackrel{\scriptstyle m,m'=0}{m\neq m'}}^{N-1} \sqrt{\frac{(a^2_m-a^2_{N-1})(a^2_{m'}-a^2_{N-1})}{[(D+1)P_\mathbbm{1}^?]^2}},\nonumber\\
&&
\label{eq:F_ME_?}
\end{eqnarray}
with $F^{\rm clas}$ given by Eq.~(\ref{eq:F_clas}). The above fidelity is lower than the optimal $F^{\rm ME}$ given by Eq.~(\ref{eq:F_opt}), and accordingly, lower than $F^{\rm MC}_{{\rm s},\mathbbm{1}}$ in Eq.~(\ref{STRATEGY-CONCLUSIVE}), as expected.  It will be, however, larger than the fidelity of teleportation via classical channel if there exist at least two Schmidt coefficients of the entangled state in (\ref{QUANTUMCHANNEL}) which differ from the minimum. 

Instead of the ME strategy, Alice could also have applied another round of MC measurements to discriminate the states $|\xi_l\rangle_2$, after an inconclusive event. This possibility and their consequences will be addressed in Sec.~\ref{sec:results2}.

\subsubsection{The overall teleportation fidelity}
\label{subsub:Overall_Fid}

The overall teleportation fidelity is computed by averaging the fidelities of conclusive and inconclusive events with their respective probabilities. Denoting this quantity by $\overline{\digamma}^\mathcal{S}$, we have for the present protocol
\begin{equation}    
\overline{\digamma}^\mathcal{S}=(1-P^?_\mathbbm{1})F^{\rm MC}_{{\rm s},\mathbbm{1}}+P^?_\mathbbm{1}\overline{\digamma}_{{\rm f},\mathbbm{1}}^\mathcal{S},
\end{equation}
where $\overline{\digamma}_{{\rm f},\mathbbm{1}}^\mathcal{S}$ represents the average fidelity when the measurement strategy (or a sequence of strategies) $\mathcal{S}$ is applied by Alice after obtaining an inconclusive result in the first stage of the MC measurement. For instance, if Alice applies the ME strategy as described above we have $\overline{\digamma}_{{\rm f},\mathbbm{1}}^\mathrm{ME}=F_{{\rm f},\mathbbm{1}}^\mathrm{ME}$. Thus, using Eqs.~(\ref{eq:min_P?}), (\ref{STRATEGY-CONCLUSIVE}), and (\ref{eq:F_ME_?}), the overall teleportation fidelity for this particular case will be given by
\begin{eqnarray}
\overline{\digamma}^{\rm ME}&=&(1-P^?_\mathbbm{1})F^{\rm MC}_{{\rm s},\mathbbm{1}}+P^?_\mathbbm{1}F_{{\rm f},\mathbbm{1}}^{\rm ME}\nonumber\\[2mm]
&=&F^{\rm clas}+Na_{N-1}^2\left(\frac{N-1}{D+1}\right)\nonumber\\[2mm]
&& \text{}+\sum_{\stackrel{\scriptstyle m,m'=0}{m\neq m'}}^{N-1} \sqrt{\frac{(a^2_m-a^2_{N-1})(a^2_{m'}-a^2_{N-1})}{(D+1)^2}}.
\label{F2}
\end{eqnarray}
This fidelity generalizes the one obtained previously in Refs.~\cite{Gu02,Roa03} for perfect conclusive teleportation, i.e., for $N=D$. Comparing it with the optimal teleportation fidelity achieved by a deterministic protocol [see Eq.~(\ref{eq:F_opt})], one can show that 
\begin{equation}     \label{eq:Overall_ME}
F^{\rm ME}\geqslant\overline{\digamma}^{\rm ME},
\end{equation}
where, again, the equality holds when the Schmidt coefficients of the channel satisfy $a_m=N^{-1/2}$ for all $m$. More generally, using the fact that the largest average fidelity of transformation of a pure bipartite state into another using LOCC is achieved  deterministically rather than probabilistically \cite{Vidal00}, we can write\footnote{When restricted to LOCC, a deterministic conversion of an initial pure bipartite state $|\Psi\rangle_{12}$ into a final pure state $|\Phi\rangle_{12}$ gives the maximal average fidelity with respect to a given target state $|\tilde\Psi\rangle_{12}$ (see lemma 2 of Ref.~\cite{Vidal00}). When the target state is a maximally entangled one, this means that the \emph{average} singlet fraction [the average of Eq.~(\ref{eq:Singlet-fraction})] achieves its maximum deterministically. Therefore, by averaging Eq.~(\ref{eq:Horodecki}) for deterministic and probabilistic protocols we arrive at Eq.~(\ref{eq:Average_fidel_general}). \label{FN:overall_Fid}} 
\begin{equation}     \label{eq:Average_fidel_general}
F^{\rm ME}\geqslant\overline{\digamma}^\mathcal{S}, 
\end{equation}
 for any $\mathcal{S}$. Therefore, the reduction of the overall teleportation fidelity in comparison with the one achieved in a deterministic protocol is another consequence of teleporting with a fidelity larger than the optimal one given by Eq.~(\ref{eq:F_opt}).

\subsection{Example: Part I}
\label{subsec:ExampleI}

To illustrate graphically what has been discussed so far, we consider the teleportation of a four-dimensional qudit state (ququart) through an arbitrary quantum channel with Schmidt rank $N=3$. In the following, we plot the fidelities as a function of the Schmidt coefficients ($a_0$ and $a_1$) of the entangled state  in Eq.~(\ref{QUANTUMCHANNEL}) without assuming any ordering between them. 

\begin{figure}[t]
\centerline{
\includegraphics[width=.5\textwidth]{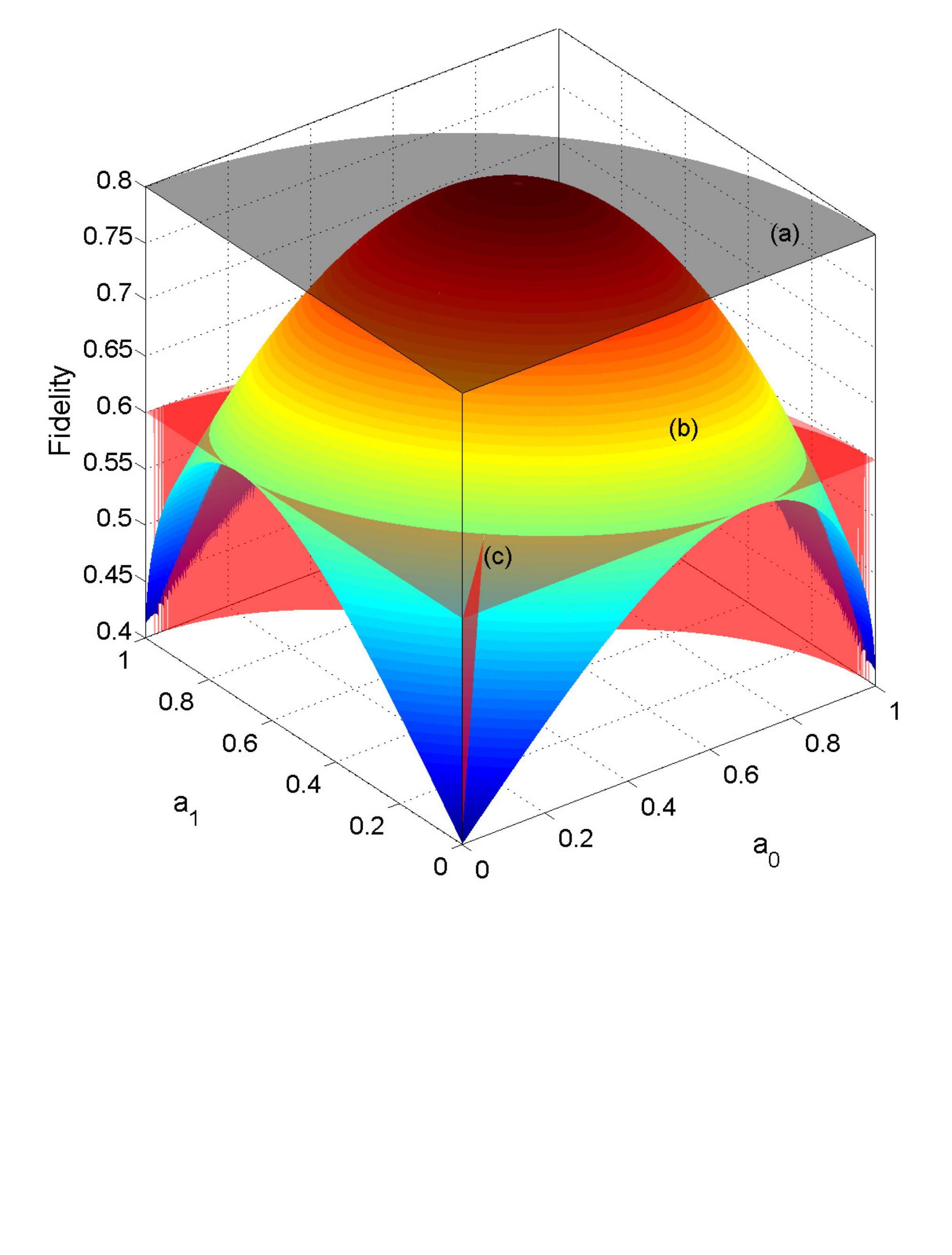}}
\vspace{-3cm}
\centerline{
\includegraphics[width=.55\textwidth]{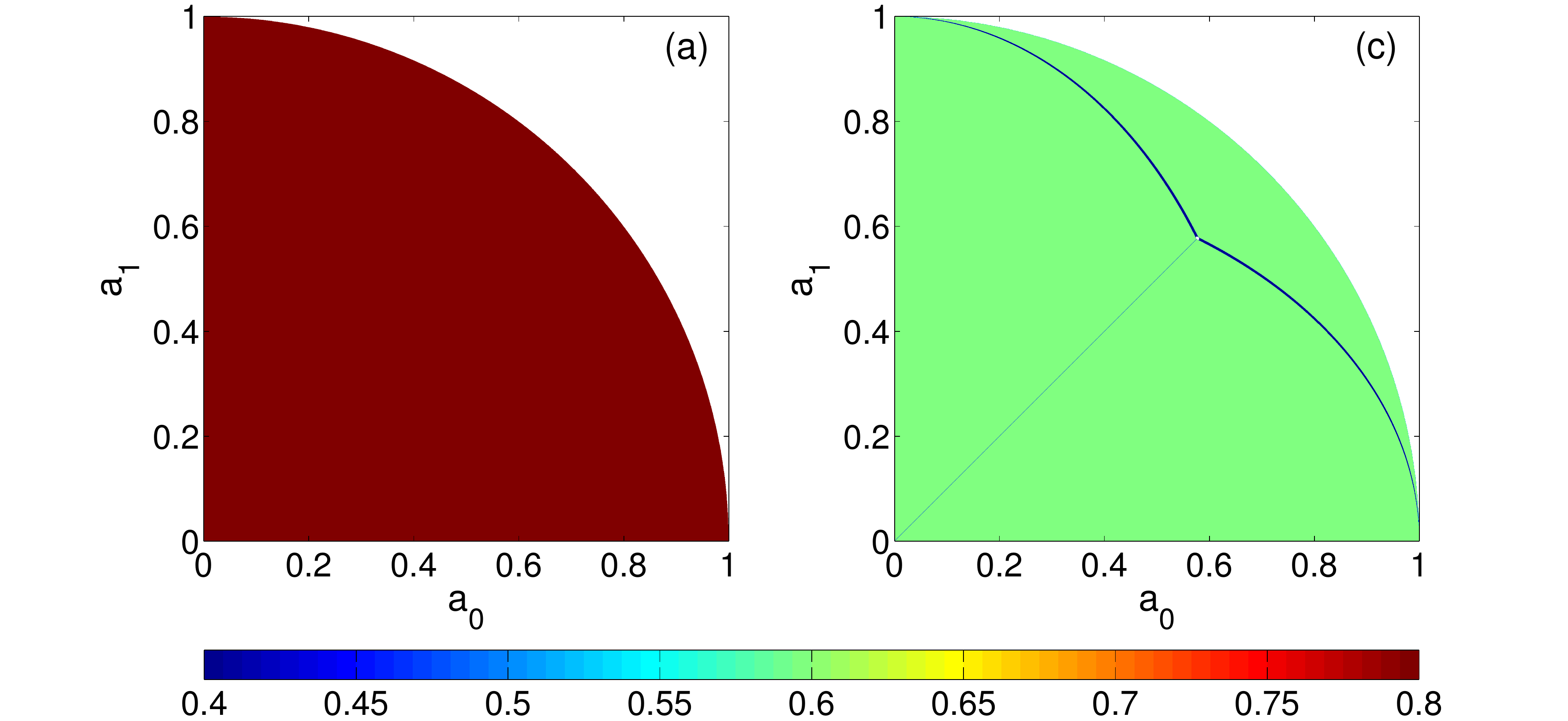}}
\caption{(Color online) Fidelity of teleportation of an unknown ququart state through an arbitrary pure quantum channel with Schmidt rank $N=3$. In the upper panel we have: (a) the conclusive fidelity $F^{\rm MC}_{{\rm s},\mathbbm{1}}$ given by Eq.~(\ref{STRATEGY-CONCLUSIVE}), (b) the optimal fidelity $F^{\rm ME}$ of a deterministic protocol given by Eq.~(\ref{eq:F_opt}), and (c) the conclusive fidelity $F^{\rm MC}_{{\rm s},\mathbbm{2}}$ given by Eq.~(\ref{f_2}). The floor of the graphic corresponds to the teleportation fidelity via classical channel, $F^{\rm clas}$, given by Eq.~(\ref{eq:F_clas}). The lower panels show the graphics (a) and (c) viewed from the top.}
\label{fig:Fidel_N3_D4}
\end{figure}

From Eq.~(\ref{STRATEGY-CONCLUSIVE}), the conclusive teleportation fidelity will be $F_{{\rm s},\mathbbm{1}}^{\rm MC}=0.8$ for any initial entangled state. This is shown in Fig.~\ref{fig:Fidel_N3_D4}(a) (upper and lower panels). The optimal fidelity achieved deterministically, $F^{\rm ME}$, and given by Eq.~(\ref{eq:F_opt}) is shown in Fig.~\ref{fig:Fidel_N3_D4}(b) (upper panel). One clearly observes the relation between those fidelities as established by Eq.~(\ref{eq:Fmc-vs-Fme}). The floor of this graphic corresponds to the maximal fidelity of teleportation via classical channel, which from Eq.~(\ref{eq:F_clas}) will be $F^{\rm clas}=0.4$.  

\begin{figure*}[t]
\centerline{
\includegraphics[width=1.2\textwidth]{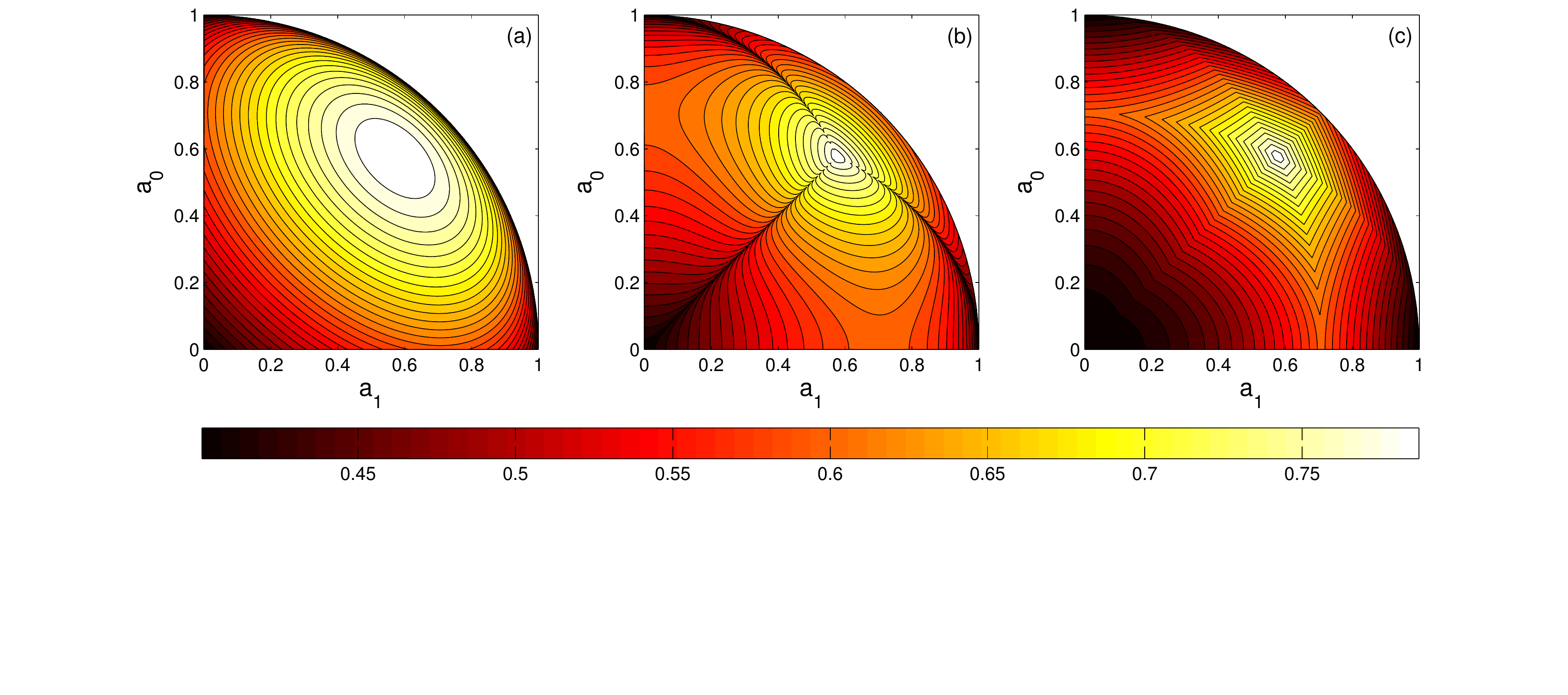}}
\vspace{-2.9cm}
\caption{(Color online) Overall fidelity of teleportation of an unknown ququart state through an arbitrary pure quantum channel with Schmidt rank $N=3$. (a) Deterministic protocol: Alice applies the ME strategy and teleports with the optimal fidelity $F^{\rm ME}$ given by Eq.~(\ref{eq:F_opt}). (b) Probabilistic protocol I: Alice first applies the MC strategy and then ME in case of an inconclusive outcome. The corresponding overall fidelity, $\overline{\digamma}^{\rm ME}$, is given by Eq.~(\ref{F2}). (c) Probabilistic protocol II: Alice applies a two-stage SMC measurement. The corresponding overall fidelity, $\overline{\digamma}^{\rm SMC}_{\mathbbm{1}\rightarrow\mathbbm{2}}$, is given by Eq.~(\ref{eq:Fid_overall_n-stage}). One can clearly observe that $F^{\rm ME}\geqslant\overline{\digamma}^{\rm ME}\geqslant\overline{\digamma}^{\rm SMC}_{\mathbbm{1}\rightarrow\mathbbm{2}}$, as established in Eq.~(\ref{eq:All_overall}).
\label{fig:Overall_fids} }
\end{figure*}

The overall teleportation fidelities are shown in Fig.~\ref{fig:Overall_fids}. In Fig.~\ref{fig:Overall_fids}(a) we plot the fidelity achieved deterministically, which is given by Eq.~(\ref{eq:F_opt}). Figure~\ref{fig:Overall_fids}(b) shows the overall fidelity given by Eq.~(\ref{F2}), which is obtained when Alice applies ME discrimination to complete the teleportation, after an inconclusive result in the first stage of the MC measurement. The relation between these two quantities established in Eq.~(\ref{eq:Overall_ME}) is clearly observed by comparing Figs.~\ref{fig:Overall_fids}(a) and \ref{fig:Overall_fids}(b).

\section{Quantum teleportation via sequential maximum-confidence measurements}
\label{sec:results2}

Following the discussion initiated in Sec.~\ref{subsub:telep_inc}, we will consider that after an inconclusive result in the first stage of the MC measurement, Alice---in agreement with Bob---keep trying to accomplish the teleportation by  discriminating the ``failure'' states $|\xi_l\rangle_2$ given by Eq.~(\ref{MCXI}). In that section we described the case where she applies the ME strategy. Here, we will describe the case where she applies another stage(s) of MC measurements. In this regard, we say that the teleportation is performed via sequential maximum-confidence (SMC) measurements \cite{Jimenez11}. Next, we briefly describe this measurement strategy and after that we discuss its application on teleportation.

\subsection{Sequential maximum-confidence measurements}
\label{subsec:SMC}

The concept of SMC measurement has been introduced in the context of discrimination among linearly dependent and equally likely symmetric pure states \cite{Jimenez11}. As we saw in Sec.~\ref{subsec:MC-review}, when this set of states is subjected to a MC measurement, an inconclusive outcome maps the $D$ input states $|\nu_l\rangle_2$ in Eq.~(\ref{SYMMETRICSTATES}) into a new set of $D$ linearly dependent and equally likely symmetric states $|\xi_l\rangle_2$ in Eq.~(\ref{MCXI}), with a one-to-one correspondence between them. For convenience we rewrite the states $|\xi_l\rangle_2$ as 
\begin{equation}     \label{eq:new_xi}
|\xi_l\rangle_2=\hat Z^l_2\sum_{k=0}^{N-\mu_1-1}\!\!\!\!b_k|k\rangle_2, \hspace{4mm} b_k=\sqrt{\frac{a^2_k-a^2_{N-1}}{P_\mathbbm{1}^?}}, 
\end{equation}
where $\mu_1$ denotes the multiplicity of the smallest coefficient of $|\nu_l\rangle_2$, namely $a_{N-1}$. Therefore, the $D$ states $|\xi_l\rangle_2$ are restricted to a $(N-\mu_1)$-dimensional Hilbert space. Since $1\leqslant\mu_1\leqslant N$, we have three possible situations: (i) If $\mu_1=N$, MC and ME strategies coincide and there is no inconclusive result, so $|\xi_l\rangle_2=0$ for all $l$. (ii) If $\mu_1=N-1$, the $D$ states $|\xi_l\rangle_2$ will be identical, up to a global phase, so no further measurement will allow us to gain information about the input states. (iii) If $\mu_1<N-1$, the MC measurement can be applied again and a conclusive outcome leads us to identify each input state with the confidence
\begin{equation}    \label{eq:MC-max-2}
[P(\nu_l|l)]_{{\rm s},\mathbbm{2}}^{\rm MC}=\frac{N-\mu_1}{D},
\end{equation}
which is smaller than the one achieved in the first stage [see Eq.~(\ref{eq:MC_probability})], as it should be. On the other hand, following Eq.~(\ref{eq:min_P?}), an inconclusive outcome occurs with the minimum probability
\begin{equation}    \label{eq:P?2}
P_{\mathbbm{2}}^?=1-(N-\mu_1)b_{N-\mu_1-1}^2,
\end{equation}
and when it occurs the states $|\xi_l\rangle_2$ are mapped to a new set of equiprobable symmetric states restricted to a ($N-\mu_1-\mu_2$)-dimensional subspace, where $\mu_2$ accounts for the multiplicity of the second smallest input-state coefficient $a_k$. Thus, the whole analysis above applies again to this new set, and the process can be iterated in as many stages as allowed by the multiplicities $\mu_j$ of the coefficients $a_k$. As we show in the Appendix, if there exist $d\in[1,N]$ sets of equal coefficients, the maximum number of conclusive stages, $\mathbbm{M}$, will be
\begin{equation}     \label{eq:Max-stages}
\mathbbm{M}=d-\delta_{\mu_d,1},
\end{equation}
where $\mu_d$ is the multiplicity of the largest $a_k$ in Eq.~(\ref{SYMMETRICSTATES}).
 This process is referred to as SMC measurement \cite{Jimenez11}.

\subsection{Teleportation via SMC measurements}

\subsubsection{Conclusive teleportation fidelities and required resources}

\begin{figure}[b]
\centerline{
\includegraphics[width=.40\textwidth]{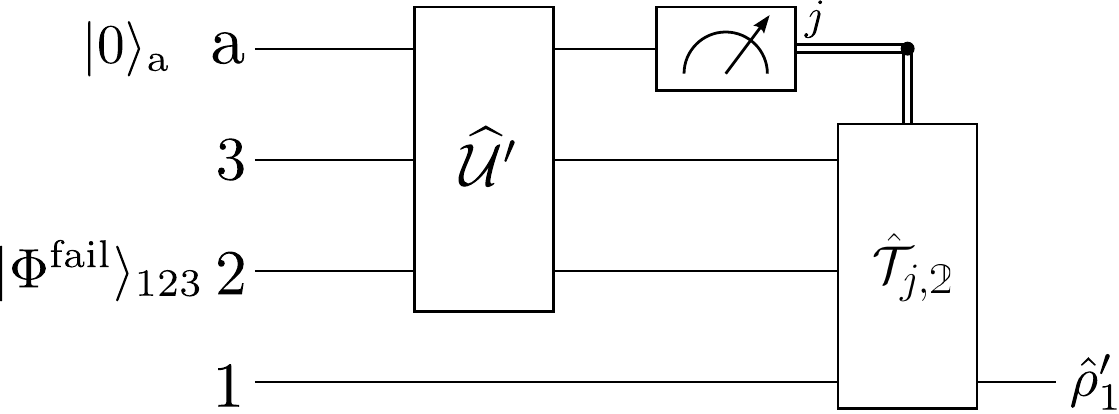}}
\caption{Quantum circuit for the second attempt of teleporting via MC measurements. It must be inserted into the box $\hat{\mathcal{T}}_{j,\mathbbm{1}}$ in Fig.~\ref{fig:Q_circuit2}(a) with $j={\rm f}$. $|\Phi^{\rm fail}\rangle_{123}$ is given by (\ref{S2}) and $\hat{\mathcal{U}}'=\hat{U}'_{2\rm a}\otimes\hat{I}_3$. The box $\hat{\mathcal{T}}_{j,\mathbbm{2}}$ represents the quantum circuit that completes the teleportation after the outcome $j$ in the measurement of the ancilla by Alice is communicated to Bob. If $j=\rm s$, $\hat{\mathcal{T}}_{\rm s,\mathbbm{2}}$ is given by the circuit of Fig.~\ref{fig:Q_circuit2}(b) and in the end the teleported state has a fidelity given by Eq.~(\ref{f_2}). If $j=\rm f$, $\hat{\mathcal{T}}_{\rm f,\mathbbm{2}}$ will depend on the strategy between Alice and Bob: a new stage of MC measurements, an inconclusive event, etc., and $\hat{\rho}'_1$ is the final state of Bob's system which depends on $\hat{\mathcal{T}}_{j,\mathbbm{2}}$. \label{fig:Q_circuit3} }
\end{figure}

Back to teleportation, let us consider now that after an inconclusive result in the first stage of the MC measurement, Alice applies this strategy again to discriminate the states $|\xi_l\rangle_2$ given by Eq.~(\ref{eq:new_xi}). For this, she needs, first, to add another two-dimensional ancilla prepared in the logical state $|0\rangle_{\rm a}$ and to implement the unitary coupling $\hat{U}'_{2\rm a}$ between it and system 2. Then she measures the ancilla on the computational basis and communicate the outcome to Bob by sending him a one-bit message through a classical channel. This process is depicted in the quantum circuit of Fig.~\ref{fig:Q_circuit3}. A conclusive outcome occurs with maximum probability $1-P_\mathbbm{2}^?$ [see Eq.~(\ref{eq:P?2})] and allows Alice and Bob to complete the teleportation process as in the standard case [e.g., see the quantum circuit of Fig.~\ref{fig:Q_circuit2}(b)]. From the discussion of Sec.~\ref{subsec:Conc_telep} and using Eq.~(\ref{STRATEGY-CONCLUSIVE}), the conclusive teleportation fidelity in the second stage of the SMC measurement will be
\begin{equation}    
F^{\rm MC}_{{\rm s},\mathbbm{2}}=\frac{N-\mu_1+1}{D+1}=F^{\rm MC}_{{\rm s},\mathbbm{1}}-\frac{\mu_1}{D+1},
\label{f_2}
\end{equation}
which is smaller than the one achieved in the first stage, as expected. However, by the same reasoning used to demonstrate Eq.~(\ref{eq:Fmc-vs-Fme}), we have that $F^{\rm MC}_{{\rm s},\mathbbm{2}}\geqslant F_{{\rm f},\mathbbm{1}}^{\rm ME}$, where $F_{{\rm f},\mathbbm{1}}^{\rm ME}$ is given by Eq.~(\ref{eq:F_ME_?}) and the equality holds when the coefficients of $|\xi_l\rangle_2$ in Eq.~(\ref{eq:new_xi}) satisfy $b_k=(N-\mu_1)^{-1/2}$ for all $k$. Thus, it is possible to achieve, with certain probability, a better teleportation fidelity by applying a MC strategy rather than ME after an inconclusive outcome in the first stage of the MC measurement.

Let us note that according to the discussion below Eq.~(\ref{eq:new_xi}) it can be shown that in order to have at least two stages of MC measurements in the teleportation protocol, the following conditions must hold simultaneously: (i) The Schmidt rank of the quantum channel in Eq.~(\ref{QUANTUMCHANNEL}) must be $N\geqslant 3$ and (ii) the multiplicity of the smallest Schmidt coefficient of this channel ($a_{N-1}$) must satisfy $\mu_1<N-1$. As we are considering teleportation of a qudit state with $D>N$, the condition (i) says that the simplest case where SMC measurements can be applied to assist the protocol is when $D=4$ and $N=3$, if condition (ii) holds. This case has been addressed in Sec.~\ref{subsec:ExampleI} and we will return to it later.

The $N$ Schmidt coefficients of the entangled state in Eq.~(\ref{QUANTUMCHANNEL}) can be grouped into $d\in[1,N]$ sets of equal coefficients. This means that Alice can apply, at most, $\mathbbm{M}$ stages of MC measurements, with $\mathbbm{M}$ given by Eq.~(\ref{eq:Max-stages}) (see also the Appendix). At each stage, the procedure is the same described above, and if she gets a conclusive outcome in the $\mathbbm{k}$th one, the teleportation fidelity, after completing the protocol, will be 
\begin{subequations}    \label{eq:Fid_sub_Eq}
\begin{eqnarray}
F_{{\rm s},\mathbbm{k}}^{\rm MC}&=&F_{{\rm s},\mathbbm{k}-1}^{\rm MC}-\frac{\mu_{\mathbbm{k}-1}}{D+1}  \label{eq:Fid_con_a} \\[2mm]
& = & \frac{1}{D+1}\left(1+\sum_{j=\mathbbm{k}}^{d}\mu_j\right),  \label{eq:Fid_con_b}
\end{eqnarray}
\end{subequations}
where $\mu_j$ denotes the multiplicity of the $j$th smallest Schmidt coefficient. Equation~(\ref{eq:Fid_con_a}) shows that  the conclusive teleportation fidelity decreases from one stage to the next, which is a consequence of the decrease in the confidence with which Alice discriminates the states of system 2 from one stage to the next \cite{Jimenez11}. To arrive at (\ref{eq:Fid_con_b}) we use that $\sum_{j=1}^{d}\mu_j=N$. Comparing this equation with (\ref{eq:Horodecki}) we find the singlet fraction
\begin{equation}     \label{eq:sing_frac_SMC}
f^{\rm MC}_{{\rm s},\mathbbm{k}}=\frac{1}{D}\sum_{j=\mathbbm{k}}^{d}\mu_j=[P(\nu_l|l)]_{{\rm s},\mathbbm{k}}^{\rm MC},
\end{equation}
where the latter equality follows from $[P(\nu_l|l)]_{{\rm s},\mathbbm{k}}^{\rm MC}=[P(\nu_l|l)]_{{\rm s},\mathbbm{k}-1}^{\rm MC}-\mu_{\mathbbm{k}-1}/D$ \cite{Jimenez11}. Finally, if Alice gets a conclusive outcome in the last allowed stage, $\mathbbm{M}$, the teleportation fidelity will be
\begin{equation}
F_{{\rm s},\mathbbm{M}}^{\rm MC}=\frac{\mu_{d-1}\delta_{\mu_d,1}+\mu_d+1}{D+1},
\end{equation}
which is always larger than the maximal one achieved via classical channel and given by Eq.~(\ref{eq:F_clas}).

To account for the required resources to accomplish teleportation via SMC measurements, let us consider that Alice and Bob agree beforehand to implement at most $\mathbbm{k}$ among the $\mathbbm{M}$ allowed stages of such strategy ($\mathbbm{k}\leqslant\mathbbm{M}$). In the worst case of successful teleportation, Alice obtains a conclusive outcome at the $\mathbbm{k}$th stage and the process will consume, in addition to the entangled state and $2\log_2(D)$ bits of classical communication, another $\mathbbm{k}$ classical bits as well as $\mathbbm{k}$ two-dimensional ancillas. On the other hand, if Alice obtains an inconclusive outcome at the $\mathbbm{k}$th stage, all the used resources are lost, i.e., $\mathbbm{k}$ classical bits and ancillas, the entangled state and the unknown qudit state. In this case, she and Bob discard the attempt and start the process again with new copies of the aforementioned resources. Therefore, this protocol is always more expensive than a deterministic one. However, as we shall see, in some situations the application of more than one stage of MC measurements enable Alice and Bob to save resources for teleportation.

\subsubsection{Overall probability of teleportation and overall fidelity}

The probability of teleportation at the $\mathbbm{k}$th stage of a SMC measurement, $P^{\rm MC}_{{\rm s},\mathbbm{k}}$, is given by
\begin{equation}     \label{eq:Prob_telep_k}
P^{\rm MC}_{{\rm s},\mathbbm{k}} =\left(1-P_\mathbbm{k}^?\right){\prod_{j=\mathbbm{1}}^{\mathbbm{k}-1}P_j^?},
\end{equation}
which takes into account the inconclusive probabilities at all $\mathbbm{k}-1$ preceding stages. Considering again that Alice and Bob agree beforehand to implement at most $\mathbbm{k}$ among the $\mathbbm{M}$ allowed stages of a SMC measurement ($\mathbbm{k}\leqslant\mathbbm{M}$), we can compute the overall probability of teleportation, $P^{\rm SMC}_{\mathbbm{1}\rightarrow\mathbbm{k}}$, just by adding the success probability at each stage. Thus, using Eq.~(\ref{eq:Prob_telep_k}), it is easy to show that 
\begin{equation}     \label{eq:Probab_Overall}
P^{\rm SMC}_{\mathbbm{1}\rightarrow\mathbbm{k}} =  \sum_{j=\mathbbm{1}}^{\mathbbm{k}}P_{{\rm s},j}^{\rm MC} = 1- \prod_{j=\mathbbm{1}}^{\mathbbm{k}}P^?_j.
\end{equation}
The overall teleportation fidelity, assuming that Alice implements a given strategy or a sequence of strategies $\mathcal{S}$ after obtaining an inconclusive outcome at the  $\mathbbm{k}$th stage, will be
\begin{equation}
\overline{\digamma}^{{\rm SMC},\mathcal{S}}_{\mathbbm{1}\rightarrow\mathbbm{k}}=\sum_{j=\mathbbm{1}}^\mathbbm{k}P_{{\rm s},j}^{\rm MC}F_{{\rm s},j}^{\rm MC} + \left(1-P^{\rm SMC}_{\mathbbm{1}\rightarrow\mathbbm{k}}\right)\overline{\digamma}^\mathcal{S}_{{\rm f},\mathbbm{k}}, 
\end{equation}
which, for $\mathbbm{k}=\mathbbm{M}$ becomes
\begin{equation}
\overline{\digamma}^{\rm SMC}_{\mathbbm{1}\rightarrow\mathbbm{M}}
=\sum_{j=\mathbbm{1}}^\mathbbm{M}P_{{\rm s},j}^{\rm MC}F_{{\rm s},j}^{\rm MC} + \delta_{\mu_d,1}\left(1-P^{\rm SMC}_{\mathbbm{1}\rightarrow\mathbbm{M}}\right)F^{\rm clas} ,
\label{eq:Fid_overall_n-stage}
\end{equation}
with $F^{\rm clas}$ given by Eq.~(\ref{eq:F_clas}). Note that the second term on the right-hand side of Eq.~(\ref{eq:Fid_overall_n-stage}) contributes only if the multiplicity, $\mu_d$, of the largest Schmidt coefficient is 1. 

According to the discussion in Sec.~\ref{subsub:Overall_Fid} and using Eqs.~(\ref{eq:Overall_ME}) and (\ref{eq:Average_fidel_general}) (see also footnote \ref{FN:overall_Fid}) we can order the overall teleportation fidelities of the protocols described so far as follows
\begin{equation}     \label{eq:All_overall}
 F^{\rm ME}\geqslant\overline{\digamma}^{\rm ME}\geqslant\overline{\digamma}^{{\rm SMC},\mathcal{S}}_{\mathbbm{1}\rightarrow\mathbbm{k}}\geqslant\overline{\digamma}^{\rm SMC}_{\mathbbm{1}\rightarrow\mathbbm{M}}.
\end{equation}
Therefore, the effect of adding further stages of MC measurements in the teleportation process is a reduction of the overall fidelity.

\subsection{Teleporting with a fidelity above \protect$F^\mathrm{ME}$}

As discussed so far, in the problem of teleportation via quantum channels with nonmaximal Schmidt rank, Alice could, in principle, apply an $\mathbbm{M}$-stage SMC measurement to assist the protocol. In this case, a conclusive event at any stage $\mathbbm{k}\leqslant\mathbbm{M}$ leads to a teleportation fidelity $F^{\rm MC}_{{\rm s},\mathbbm{k}}$ given by Eq.~(\ref{eq:Fid_sub_Eq}). On the other hand, Alice---in agreement with Bob---could have chosen to apply a ME measurement and teleport deterministically with an optimal fidelity $F^{\rm ME}$ given by Eq.~(\ref{eq:F_opt}). However, let us assume that for Alice and Bob purposes, it is sufficient to accomplish teleportation with a fidelity of transmission better than a threshold set by $F^{\rm ME}$. If their shared quantum channel enables Alice to implement an $\mathbbm{M}$-stage SMC measurement, each stage $\mathbbm{k}$ whose conclusive fidelity fulfills this condition will be useful. This is always the case for the first stage (unless $a_m=N^{-1/2}$ for all $m$), as shown in Eq.~(\ref{eq:Fmc-vs-Fme}). For the remaining stages the condition $F^{\rm MC}_{{\rm s},\mathbbm{k}}> F^{\rm ME}$, obtained from Eqs.~(\ref{eq:F_opt}) and (\ref{eq:Fid_sub_Eq}), reads\footnote{Within the framework of quantum state discrimination, we can use Eqs.~(\ref{eq:sing_frac_ME}) and (\ref{eq:sing_frac_SMC}) to write the inequality (\ref{eq:ineq}) as
$[P(\nu_l|l)]_{{\rm s},\mathbbm{k}}^{\rm MC}>[P(\nu_l|l)]^{\rm ME}.$
This expression says that the $\mathbbm{k}$th stage of a SMC measurement will be useful for teleporting with a fidelity above $F^{\rm ME}$, whenever a conclusive outcome in that stage makes Alice's confidence in identifying the states $|\nu_l\rangle_2$ [see Eq.~(\ref{SYMMETRICSTATES})] larger than the one achieved via the ME strategy. }
\begin{equation}     \label{eq:ineq}
\sum_{j=\mathbbm{k}}^{d}\mu_j> \left(\sum_{m=0}^{N-1} a_m\right)^2.
\end{equation}
Given an arbitrary entangled state as in Eq.~(\ref{QUANTUMCHANNEL}), it is not possible to obtain a general expression for the maximum number of useful stages among the $\mathbbm{M}$ allowed ones [see Eq.~(\ref{eq:Max-stages})] in a SMC measurement. However, since Alice and Bob have complete knowledge of this state, all the parameters in the above inequality are known beforehand. Thus, before starting the teleportation, they have to verify whether the $\mathbbm{k}$th stage ($1<\mathbbm{k}\leqslant\mathbbm{M}$) will be useful and then establish the number of stages to be implemented in the protocol. 

Assuming that a $\mathbbm{k}$-stage SMC measurement enables Alice and Bob to accomplish teleportation with a fidelity  above $F^{\rm ME}$, the overall probability of success in this case will be given by Eq.~(\ref{eq:Probab_Overall}). From this equation, we clearly have
\begin{equation}
P^{\rm SMC}_{\mathbbm{1}\rightarrow\mathbbm{k}}\geqslant P_{{\rm s},\mathbbm{1}}^{\rm MC},
\end{equation}
where the equality holds for $\mathbbm{k}=\mathbbm{1}$. Therefore, the implementation of further $\mathbbm{k}-1$ stages, in addition to the first one, increases the chance of teleportation. This is specially important in a scenario where Alice and Bob have limited resources, mainly those considered to be more expensive, as for instance, the entangled states. If so, instead of simply discarding the teleportation attempt after an inconclusive outcome in the first stage, they can keep trying to teleport in one of the remaining $\mathbbm{k}-1$ stages. This increases the overall probability of teleportation with a fidelity above the threshold $F^{\rm ME}$ and, consequently, allows them to save resources.

\subsection{Example: Part II}

We return now to the example of teleportation of an unknown ququart state through an arbitrary quantum channel with Schmidt rank $N=3$. As mentioned earlier, this is the simplest case where teleportation via SMC measurements may be applied. Here, we have at most two stages, the first of which has been discussed in Sec.~\ref{subsec:ExampleI}. If an inconclusive outcome in that stage is followed by a conclusive one in the second, the teleportation fidelity, given by Eq.~(\ref{f_2}), will be $F^{\rm MC}_{{\rm s},\mathbbm{2}}=0.8-\mu_1/5$. This is shown in Fig.~\ref{fig:Fidel_N3_D4}(c) (upper and lower panels). In those graphics, the lines where the fidelity is equal to the classical one represents the entangled states whose smallest Schmidt coefficient has multiplicity $\mu_1=2$. The point where the three lines touch each other represents an event that never occurs because $\mu_1=3$, and, hence, there is no inconclusive outcome at the first stage. For the points on the surface \ref{fig:Fidel_N3_D4}(b) that are below the plateau in Fig.~\ref{fig:Fidel_N3_D4}(c), the corresponding entangled states enable conclusive fidelities above $F^{\rm ME}$ at the second stage of the SMC measurement. Figure~\ref{fig:Overall_fids}(c) shows the overall teleportation fidelity, $\overline{\digamma}^{\rm SMC}_{\mathbbm{1}\rightarrow\mathbbm{2}}$, obtained from Eq.~(\ref{eq:Fid_overall_n-stage}). Comparing this graphic with those from Figs.~\ref{fig:Overall_fids}(a) and \ref{fig:Overall_fids}(b), one can verify the relation established in Eq.~(\ref{eq:All_overall}).

\begin{figure}[t]
\centerline{
\includegraphics[width=.55\textwidth]{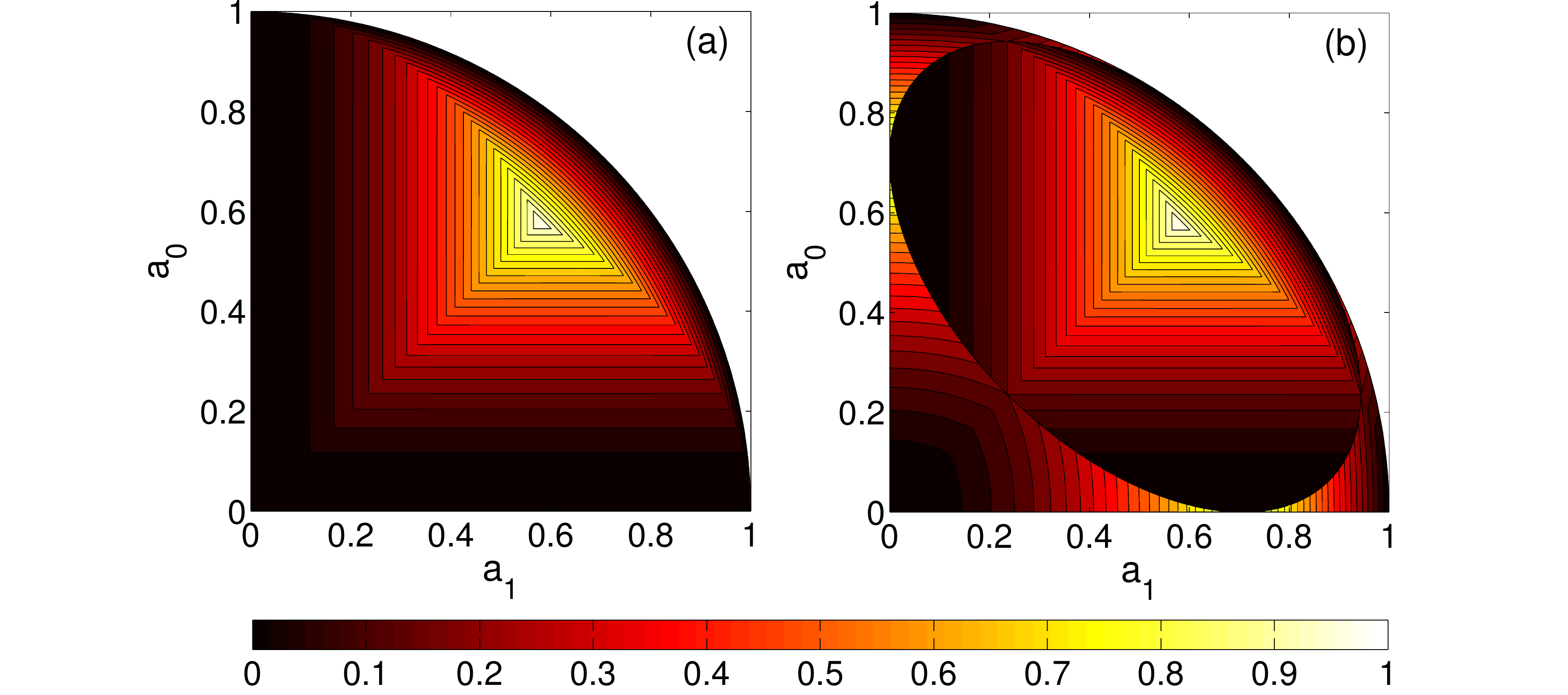}}
\caption{(Color online) Probability of teleportation of an unknown ququart state through an arbitrary pure quantum channel with Schmidt rank $N=3$. (a) Probability at the first stage of the SMC measurement given by $1-P^?_\mathbbm{1}$ [see Eq.~(\ref{eq:min_P?})]. (b) Overall probability of teleportation [see Eq.~(\ref{eq:Probab_Overall})] with a fidelity above $F^{\rm ME}$. }
\label{fig:Prob_N3}
\end{figure}

In Fig.~\ref{fig:Prob_N3}(a) we plot the probability of teleportation at the first stage of the SMC measurement, obtained from Eq.~(\ref{eq:min_P?}). With those probabilities, we have a teleportation fidelity above $F^{\rm ME}$ for any entangled state [see Eq.~(\ref{eq:Fmc-vs-Fme})]. Figure~\ref{fig:Prob_N3}(b) shows the \emph{overall} probability of teleportation  [see Eq.~(\ref{eq:Probab_Overall})] with the condition that the conclusive fidelity of transmission be larger than $F^{\rm ME}$. The region in this graphic where the probabilities exceeds those ones in Fig.~\ref{fig:Prob_N3}(a) represents the entangled states where, in addition to the first, the second stage of the SMC measurement also satisfy the condition. For these states the implementation of a $\mathbbm{2}$-stage SMC measurement is clearly advantageous.

\section{Summary and conclusion}
\label{summary}

To summarize, in this work we have described the quantum teleportation protocol via nonmaximally entangled pure states within the framework of quantum state discrimination. More specifically, we showed that a crucial step for teleportation is related to the ability of the sending party, Alice, to discriminate nonorthogonal symmetric states which are equally likely. The type of teleportation that she and Bob (the receiving party) prearrange to implement defines the discrimination strategy that she has to adopt. In this regard, we have shown that to achieve optimal teleportation fidelity in a deterministic way \cite{Vidal00,Banaszek00,Banaszek01,Laustsen02}, Alice must implement an optimized ME strategy, which is valid for both linearly dependent and linearly independent symmetric states. On the other hand, to achieve unit teleportation fidelity with maximum success probability, Alice has to apply an optimized UD strategy. This is a known fact since this protocol, called perfect conclusive teleportation, has been conceived \cite{Mor96} and intensively studied thereafter \cite{Li00,Bandyopadhyay00,Son01,Gu02,Roa03}. However, UD applies only for linearly independent states \cite{Chefles98-2,Chefles98-1}, which implies that perfect conclusive teleportation is restricted to quantum channels with maximal Schmidt rank. Here, we investigated the problem where this latter condition does not hold. It was shown that if Alice applies an optimized MC measurement \cite{Croke06} to assist teleportation in this case, the fidelity of transmission at conclusive events, although less than unit, is the maximum achievable for that channel. In fact, as can be seen from Eqs.~(\ref{STRATEGY-CONCLUSIVE}) and (\ref{eq:Fmc-vs-Fme}), we found that this fidelity depends only on the Schmidt rank of the channel and it is better than the one achieved deterministically via ME measurements. Moreover, the probability of successfully teleporting is also maximal. This optimal imperfect conclusive teleportation was shown to interpolate between the optimal deterministic (when the Schmidt coefficients are all equal) and the perfect conclusive (when the Schmidt rank is maximal) protocols.

Subsequently, the teleportation via SMC measurements, introduced in Ref.~\cite{Jimenez11}, has also been investigated. In this strategy, one takes into account that it might be possible to extract some information about the input states even after an inconclusive result in the MC measurement. If so, MC can be applied again and the process can be iterated until no further information is available. We have shown that if the quantum channel allows for the maximum of $\mathbbm{M}$ conclusive stages in the SMC measurement, the teleportation fidelity decreases from one stage to the next as can be seen in Eq.~(\ref{eq:Fid_con_a}); at the $\mathbbm{M}$th stage, it is still always larger than the maximal fidelity achieved via classical channel. An important feature of this scheme is that there are quantum channels for which it is possible to implement $\mathbbm{k}$ among the $\mathbbm{M}$ allowed stages in the SMC measurement such that a conclusive event at any stage leads to a teleportation fidelity above the optimal one achieved deterministically. As a consequence, the overall probability of teleportation increases, allowing Alice and Bob to save the resources used in the protocol.  It is worth noting that SMC measurements could also be applied after a failed attempt of perfect conclusive teleportation, since in this case the states to be discriminated becomes linearly dependent and cannot be unambiguously identified by any further process. 

In this work we have shown that the singlet fraction of a \emph{pure} bipartite quantum channel, defined by Eq.~(\ref{eq:Singlet-fraction}), is equal to the maximum confidence with which Alice can discriminate equally likely symmetric pure states by applying a given optimized measurement strategy. This is a direct consequence of the connection between singlet fraction and entanglement concentration procedures \cite{Horodecki99,Vidal00}, and the connection between the latter and state discrimination \cite{Chefles98-2,Yang09,Croke08}. It may be fruitful, somehow, to look at the singlet fraction by this new perspective and, for instance, to extend our study here for teleportation via mixed entangled states. Finally, another possible extension of the results obtained in this work is to investigate other quantum communication protocols performed via quantum channels with nonmaximal Schmidt rank. For instance, perfect conclusive entanglement swapping via UD strategy has been studied in \cite{Delgado05}, and in a forthcoming paper we shall generalize this for an imperfect protocol accomplished via MC measurements.

\begin{acknowledgments}
This work was supported by CONICYT PFB08-24, Milenio ICM P10-030-F, PDA25, and FONDECYT Grant No. 1080383. M.A.S.P. acknowledges financial support from CONICYT.
\end{acknowledgments}

\appendix*
\section{}
\label{apend:SMC}

The maximum number of conclusive stages allowed in the SMC measurement, i.e., the number of stages in which it is possible to  gain some information about the input states through a measurement, ranges from 1 to $N-1$. It depends on the multiplicities of the input-state coefficients $\{a_k\}$ as follows. The $N$ coefficients $a_k$ can be grouped into $d$ sets of equal coefficients, where $d\in[1,N]$. The number of elements in each set gives the multiplicity of the coefficient. If $\mu_j$ (with $j=1,\ldots,d$) is the multiplicity of the $j$th smallest coefficient, then $\mu_d$ will be the multiplicity of the largest one. As we saw in Sec.~\ref{subsec:SMC}, after each stage of the SMC measurement, the dimension of the subspace where the states to be discriminated are restricted, decreases. If there are $d$ sets of equal input-state coefficients, the dimension of the Hilbert space at each stage will be
\begin{equation}     \label{eq:Diagram_Ap}
\begin{array}{ccccl}
1^{\rm st}\ \text{stage} &&\rightarrow&& N, \\[2mm]
2^{\rm nd}\ \text{stage} &&\rightarrow&& N-\mu_1, \\[2mm]
3^{\rm rd}\ \text{stage} &&\rightarrow&& N-(\mu_1+\mu_2), \\[2mm]
\vdots  &&&& \hspace{8mm}\vdots \\[2mm]
d^{\rm th}\ \text{stage} &&\rightarrow&& N-\displaystyle\sum_{j=1}^{d-1}\mu_j=\mu_d. 
\end{array}
\end{equation}
We note that if $\mu_d=1$ the failure states in the last stage will be identical, up to a global phase, so no further measurement allows us to gain information about the input states. Otherwise, if $\mu_d>1$ the failure states can be discriminated for the last time. In this case, MC and ME strategies coincide and there is no further inconclusive result. Therefore, the maximum number of conclusive stages in the SMC measurement, denoted by $\mathbbm{M}$, will be
\begin{equation}
\begin{array}{ccccc}
\mathbbm{M}=d-1, && \text{if} && \mu_d=1, \\[2mm]
\mathbbm{M}=d, && \text{if} && \mu_d>1,
\end{array}
\end{equation}
which, for short, can be written as $\mathbbm{M}=d-\delta_{\mu_d,1}$. In the extreme cases mentioned above, this will be 1 when $d=1$ (i.e., all coefficients $a_k$ are equal) and $N-1$ when $d=N$ (i.e., all coefficients $a_k$ differ).

\end{document}